\documentstyle[11pt,fleqn,leqno]{article}
\begin{document}

\title{A demonstration that the observed neutrinos
are not Majorana particles}
%\footnote{Whatever it is, I'm against it. (Groucho Marx)} 
\author{	
        R.Plaga\thanks
        {plaga@hegra1.mppmu.mpg.de}
       \\Max-Planck-Institut f\"ur Physik(Werner-Heisenberg-Institut)
        \\ F\"ohringer Ring 6
        \\D-80805 M\"unchen, Germany
	}
\date{}
\maketitle
%\centerline{format is PRL}
\begin{abstract}
\noindent
It is shown that Majorana neutrinos cannot couple
vectorially 
to the neutral-current SU(2)$_{L}$ x U(1) gauge field of
the standard model. 
Since strong evidence for the existence of such a vector coupling
in neutral current reactions
has recently been presented by the Charm II collaboration, 
it is unlikely that  the observed neutrinos
are predominantly Majorana particles.
Theorems on the ``reappearance'' of vector interactions in neutral
current scattering of Majorana neutrinos and the indistinguishability
of Majorana and Dirac neutrinos in the massless case are
discussed critically.
\end{abstract}
PACS number(s):  14.60.S,13.15,12.15.M
%Neutrinos
% -non-standard-model, 14.60.S (ordinary without S)
% -interactions, 13.15
% -mass and mixing, 14.60.P
% Neutral currents, 12.15.M
%Leptons
% -properties, 14.60
%Majorana-Weyl fields, 04.50
%\vskip 2in
%\tableofcontents    
%\include{introduction}
%\include{diracn11}
%\include{sep1pic}
\section{Introduction}
The answer to the question of whether the neutrino is a Dirac
or a Majorana particle is considered to be one of
the most important clues to physics
beyond the standard model.
If the neutrino were a ``Majorana particle''
(i.e. a particle identical to its antiparticle\cite{majorana}),
the so called ``see-saw'' mechanism could naturally
explain the smallness of neutrino masses,
which remains puzzling
within the standard model
(see ref.\cite{bil} for a recent review on
neutrino masses).
The ``see-saw'' mechanism requires the existence of
Higgs-field configurations 
beyond the one of the standard model,
which makes the prospect of experimentally proving
the Majorana nature of neutrinos (e.g. via
neutrinoless double beta-decay 
experiments) very attractive.
\\
In this {\sl paper} I present a contribution
to this long-standing problem, concluding that the
neutrino species observed up to now cannot be predominantly
Majorana particles.
If neutrinos
have Dirac character, the existing neutral-current scattering
data are in complete agreement with 
the standard model of particle physics.
In particular the neutral-current vector coupling
of neutrino, contained in the standard model, is necessary
for a satisfactory description of the experimental data
(section 3 and appendix 4, second part).
In section 2 it is shown that for Majorana particles 
any neutral-current vector coupling is forbidden. Therefore 
the experimental data cannot be quantitatively understood
in the standard way under the assumption
that neutrinos are Majorana particles.
It will probably still be possible to call in some new physics,  
which is fine tuned to explain the experimental data under
the assumption that neutrinos are Majorana particles.
Such new physics does not seem to be
required in a natural way, though (e.g. by the existence of
Higgs-field configurations leading to Majorana masses).
Faced with this situation it seems quite likely that the 
known neutrino species are Dirac particles.
I will confine the demonstration to purely neutral-current reactions.
\\
The ``modern'' choice for the metric
of the 4-vectors (defined e.g. in the textbooks of
Bjorken and Drell\cite{bjorken} and 
Mandl and Shaw\cite{mandl}) is used.
If not otherwise noted  (e.g.  in eq.(\ref{weyl})
I will work in the Majorana representation\cite{majorana,mandl}
\footnote{see the appendix of Mandl and Shaw's book\cite{mandl}
of the explicit presentation of the 
$\gamma$ matrices as used in this {\sl paper},
all five $\gamma$ matrices are purely imaginary.} 
for the $\gamma$ matrices in the
Dirac equation
(Pauli's fundamental theorem states that the choice of the
representation can have no influence on any physical
result of the theory\cite{sakurai}).
The space-time 
arguments of all fields are taken as positive.
The discussion will be in the q-number 
formalism throughout (full second-quantized field theory).
\section{The vanishing of the vector coupling in the 
Dirac equation for Majorana fields}
The general Dirac equation for a complex valued neutrino 
field operator (operators are symbolized by the 
hat $\hat{ }$) $\hat{\Psi}_{\nu}$ 
of arbitrary helicity and rest mass m$_{\nu}$
in a neutral weak boson field
Z$_{\mu}$, is obtained from the standard model
Langrangian\cite{mandl} via the Euler-Lagrange equations as:
\begin{equation}
i  \gamma^{\mu} \left( \hbar{{\partial}\over{\partial x^{\mu}}}
- {ie\over 2\cos(\theta_w) \sin(\theta_w) c} Z_{\mu} 
(g_V^{\nu}- g_A^{\nu}\gamma^{5}) \right)
\hat{\Psi}_{\nu} - m_{\nu} c \hat{\Psi}_{\nu} = 0
\label{nel}
\end{equation}
here $e$ is the positron charge, $\theta_W$
is the Weinberg angle and $g^{\nu}_V$= $g^{\nu}_A$=1/2 are
the vector and axial couplings of the neutrino
to the Z$_{\mu}$ field. $\hat{\Psi}_{\nu}$ is a Dirac bispinor.
The aim is now to find the corresponding equation
of motion
for the ``abbreviated''\cite{pauli} case of a Majorana neutrino.
\\
A Majorana particle (symbolized by the subscript M) 
is defined by the ``supplementary condition'' 
that the field and its charge conjugate 
(symbolized by the superscript c)
are identical\cite{majorana,pauli}
for all positions $\vec{x}$,t in space time:
\begin{equation}
\hat{\Psi}_M^{c}(\vec{x},t) = \hat{\Psi}_M(\vec{x},t)
\label{sc}
\end{equation}
It is possible to introduce a purely conventional phase factor
in the definition of this condition. 
I follow a usual practice 
(and Majorana's original publication\cite{majorana})
and set this factor to 1. 
\\
``Charge conjugation'' is defined as taking the
hermitian conjugate of the field operator and multiplying it
with a ``charge conjugation matrix'' S$_C$ 
\footnote{
In order to avoid confusion I use Sakurais'
symbol ``S$_C$'' for the charge 
conjugation matrix in (eq.(\ref{majc})) rather than the
more usual ``C''.
Many authors define 
C=S$_C \gamma^0$\cite{bjorken,sakurai} while others
use the notation C = S$_C$\cite{pauli}.
In spite of these differences in notation the definition of a 
``Majorana particle'' (eq.(\ref{sc})) is unequivocal.}
which is defined
by the condition 
S$_C^{-1}$ $\gamma^{\mu}$S$_C$=-$\gamma^{{\mu}*}$
\cite{sakurai} hence:
\begin{equation}
 \hat{\Psi}^c(\vec{x},t)=
S_C \hat{\Psi}^{\dag T}(\vec{x},t).
\label{majc}
\end{equation}
Here the transpose operation T only brings the bispinor
back to a column form and does not otherwise act
on the operator\cite{sakurai}.
In the Majorana representation S$_C$
 is the unit matrix I \cite{pauli,sakurai},
charge conjugation is equivalent to hermitian conjugation
and the Majorana field is necessarily\footnote{
Racah writes (my translation
from the Italian) \cite{racah}: ``The imposition of real valuedness
on the neutrino wavefunction ... is a logical consequence
of the hypothesized physical identity of neutrinos and
antineutrinos.''}
real valued\cite{majorana}. 
The first appendix gives a more mathematical explanation
of this ``real valuedness'' in the field case.
\\
According to equation (\ref{sc}) 
one can represent a field fulfilling
condition (\ref{sc}) by demanding that it is a superposition
of a Dirac field $\hat{\Psi}$ 
and its charge conjugate for all $\vec{x}$,t \cite{pauli}:
\begin{equation} 
\hat{\Psi}_M(\vec{x},t)={1 \over \sqrt{2}}(\hat{\Psi}^c(\vec{x},t)
+\hat{\Psi}(\vec{x},t))
\label{majco}
\end{equation}
In the Majorana representation the equation of motion 
for the charge conjugate field
of the neutrino
$\hat{\Psi}_{\nu}^c$ then simply follows by taking
the hermitian conjugate of  
eq.(\ref{nel})
(as the $\gamma$ matrices 
are purely imaginary in the Majorana representation
they change sign under hermitian conjugation):
\begin{equation}
i  \gamma^{\mu} \left( \hbar{{\partial}\over{\partial x^{\mu}}}
+ {ie\over 2\cos(\theta_w) \sin(\theta_w) c} Z_{\mu} 
(g_V^{\nu}+g_A^{\nu}\gamma^{5}) \right) 
\hat{\Psi}_{\nu}^c - m_{\nu} c \hat{\Psi}_{\nu}^c = 0
\label{npos}
\end{equation}
We obtain the equation of motion for a Majorana neutrino
by adding 
eq.(\ref{nel}) and eq.(\ref{npos}) and identifying
$\hat{\Psi}_M$ in the sum according to
eq.(\ref{majco}):
\begin{equation}
i  \gamma^{\mu} \left( \hbar{{\partial}\over{\partial x^{\mu}}}
+ {ie\over 2\cos(\theta_w) \sin(\theta_w) c } Z_{\mu}
g_A^{\nu} \gamma^{5} \right) 
\hat{\Psi}_{M\nu} - m_{\nu} c \hat{\Psi}_{M\nu} = 0
\label{nmaj}
\end{equation}
This equation of motion is equivalent to the one for
a Dirac neutrino (eq.(\ref{nel})) with $g_V^{\nu}$=0
(vanishing of neutrino vector coupling).
The fact that ``Majorana projections''(eq.(\ref{majco}))
only ``persist in time'' (i.e. fulfill equation (\ref{sc}) for all
t) if they do not couple via vector interactions was 
already pointed out immediately after Majorana's
original work by Furry\cite{furry}. 
He also noted that scalar interactions are possible
for Majorana neutrinos.
We now recognize
that axial coupling (not mentioned by Furry) is also allowed.
From the study of the phenomenology of supersymmetric
particles it is already known that
vector couplings have to be absent in general
for all Majorana fields $\lambda$\cite{haberkane} i.e.:
\begin{equation}
\bar{\lambda} \gamma^{\mu} \lambda = 0
\label{vec}
\end{equation}
\section{The experimental data on neutral-current elastic 
neutrino-electron scattering}
Recent experiments on the neutral-current coupling of neutrinos
show that eq.(\ref{nmaj}) does not properly describe
the observed neutrinos.
In its experiment on the purely neutral-current 
scattering of muon neutrinos on electrons
the Charm II collaboration found
for the effective neutral-current coupling constant\cite{charm2}:
$g_V^{\nu e}$ =
-0.035$\pm$0.017 (combined statistical and systematical error).
The effective coupling constant is
given as\cite{charm2}:\footnote{
For a more detailed explanation of this equation and 
the conclusion of $g_V^{\nu_{\mu}} \neq 0$ from
the Charm II data, see the second part of appendix 4.} 
\begin{equation}
g_V^{\nu e} = 2 g_V^{\nu_{\mu}} \cdot g_V^{e}
\label{coup}
\end{equation}
where $g_V^{e}$ is the vector-coupling constant of the electron
to the Z$_{\mu}$ field.
For Majorana neutrinos
from eq.(\ref{coup}) and $g_V^{\nu_{\mu}}$=0 (eq.(\ref{nmaj}))
we would expect $g_V^{\nu e}$=0 for Majorana
neutrinos which is more than
two sigmas away from the measured value.
The measured value for $g_V^{\nu e}$ is in excellent
agreement with the assumption of 
standard model values for the vector coupling constant
of the electron ($g_V^{e}$=-0.037$\pm$0.0006 \cite{ppb}) 
and a Dirac neutrino ($g_V^{\nu_{\mu}}$=1/2\cite{mandl}).
For the neutral-current effective axial coupling Charm II found
$g_A^{\nu e}$ =
-0.503$\pm$0.017 which is
consistent with the standard model expectation
($g_A^{\nu e}$=-0.507$\pm$0.0004 \cite{ppb})
for both eq.(\ref{nel}) and eq.(\ref{nmaj})
\footnote{
The method to select the quoted solution 
for their result
used by the Charm collaboration
(based on e$^+$-e$^-$ data)
has to be disregarded in our case, because 
it implicitly assumes standard model values
for the coupling constants of the neutrino.
The other three 
possible solutions in g$_V^{\nu e}$,g$_A^{\nu e}$
for the neutrino scattering results 
found by the collaboration are, however, also in
disagreement with the values expected for
Majorana neutrinos.}. 
The result for vector coupling
disfavors the identification of the muon neutrino
(and by analogy also the other neutrino flavors)
as a Majorana particle at the $>$ 95 $\%$ confidence level.
\\
That the observed neutrinos are not Majorana particles,
is not in conflict with
previous work on Majorana neutrinos (masses, mixing,
see-saw etc.). These ideas could still apply either
to a small admixture to the known neutrinos or a new
species of neutrino (for example
a heavy fourth generation neutrino\cite{turner}).
Imposing the ``Majorana supplementary condition'' (\ref{sc}) is quite
reasonable and can be physically ``explained'' e.g. by
a ``see-saw'' mechanism. It necessarily
leads to particles with no vector coupling, however.
The properties of Majorana neutrinos
thus remain a fascinating topic for further research. 
\section{Criticism of the
Kayser/Shrock argument on the vector 
coupling in the neutral currrent}
The fact that the vector part of the neutrino
current vanishes for Majorana neutrinos,
thus leading to a different
neutral-current scattering cross section
for Majorana as compared to Dirac
neutrinos, had already been clearly stated
by Kayser and Shrock\cite{kaysershrock,kayser},
who drew a different conclusion than the present paper, though.
Their argument can be summarized as follows:
\\
``In spite of the absence of vector
coupling in the interaction Langrangian for Majorana
neutrinos the vector interaction
``reappears'' because the ``empirically observed''
highly relativistic
neutrino is a ``left-handed'' state. The neutrino spinor can
thus be multiplied by a ``state preparation factor''
P$_L$=(1-$\gamma^5$)/2 without changing it:
\begin{equation}
\hat{\Psi}_{L{\nu}} = P_L \hat{\Psi}_{L{\nu}}
\label{sub}
\end{equation}
If one performs this substitution for $\hat{\Psi}_{M \nu}$
in the axial interaction term of eq.(\ref{nmaj}) the vector part
of the interaction is recovered. Therefore
Majorana and Dirac neutrinos have the same
neutral-current interaction in principle.'' (end of 
my summary of the Kayser/Shrock 
argument).
\\
Though formally correct,
there must be some logical fallacy in this reasoning:
one finds that a given special state
of the neutrino (namely a chiral left-handed one,
i.e. with chirality=-1)
leads to vector parts in the interaction Langrangian
in direct contradiction with the original
eq.(\ref{nmaj}) and a general 
theorem of Majorana fermions (eq.(\ref{vec})).
The conclusion can then only be that
this state (whether  experimentally observed or not) cannot
occur for Majorana fermions.
\\
States of {\it chirality}=-1 are {\bf indeed} forbidden
for Majorana neutrinos: charge conjugation
as defined by eq.(\ref{majc})
turns chiral left-handed states into right-handed ones,
which is in contradiction with the mathematical identity 
required by eq.(\ref{sc}) for Majorana fermions 
(see appendix 2 for a more detailed discussion).
\\
States of {\it helicity}=-1 are not necessarily in contradiction
with eq.(\ref{sc}) (appendix 3).
It is therefore not possible to exclude an identification of
the observed neutrinos
as Majorana fermions merely by way of their empirically
proven ``left-handedness''. 
However, Majorana states with helicity=-1 {\bf cannot}
fulfill eq.(\ref{sub}) (appendix 3), as erronously assumed
in the argument of Ref.\cite{kaysershrock}.

\section{The distinction between Lee-Yang and Majorana fields
for vanishing rest-mass}
There is a widely held
conviction that the Lee-Yang two-component neutrino
theory 
is equivalent to the Majorana 
abbreviation for the case 
of m$_{\nu}$=0 (``Dirac-Majorana Confusion
Theorem''\cite{case,kayser}).
I disagree in the following sense:
the Lee-Yang neutrino\cite{leeyang}
(i.e. a massless Dirac neutrino interacting
via V-A coupling) and the Majorana neutrino 
are both
``two-component neutrinos''. In spite of this
fact these cases are physically distinguishable because
eq.(\ref{nel}) and eq.(\ref{nmaj}) remain different also
for the case m$_{\nu}$=0, due to
the presence
of vector coupling in eq.(\ref{nel}). These two possibilities
for ``two-component'' neutrinos are now
examined in further detail.
%\pagebreak
\\
$\bullet$ In the Weyl representation
(denoted by the superscript ``W'')
eq.(\ref{nel}) can be written as the following system
of two equations\cite{sakurai}:
\begin{eqnarray} i \hbar \left({{\partial}\over{\partial x^{0}}} +
{{i k(g_V^{\nu} - g_A^{\nu})} \over {\hbar c}}Z_0-
\vec{\sigma} \cdot \nabla -  {{i k(g_V^{\nu} - g_A^{\nu})}
\over {\hbar c} }
\vec{\sigma} \cdot \vec{Z}
\right) \hat{\Psi}^W_R -m_{\nu} c \hat{\Psi}^W_L = 0
\nonumber
\\
i \hbar \left( {{\partial}\over{\partial x^{0}}} - 
{{i k(g_V^{\nu} +g_A^{\nu})} \over {\hbar c}}Z_0+
\vec{\sigma} \cdot \nabla + {{i k(g_V^{\nu} +g_A^{\nu})}
\over {\hbar c} }
\vec{\sigma} \cdot \vec{Z}
\right) \hat{\Psi}^W_L -m_{\nu} c \hat{\Psi}^W_R = 0
\label{weyl}
\end{eqnarray}
where $k$=$e$/( 2 cos($\theta_w$) sin($\theta_w$)) and
$\vec{\sigma}$ is the 3-vector of the Pauli matrices
in standard form.
The Lee-Yang neutrino 
(a special case of Weyl's massless two-component fermion
\cite{weyl}), can be 
described by the equations (\ref{weyl})
for the case of vanishing rest mass m$_{\nu}$. In this case
the two equations decouple
and the observed neutrinos can be fully described 
by the chiral left handed  field 
$\hat{\Psi}^W_L$ fulfilling the upper equation of (\ref{weyl}).
$\hat{\Psi}^W_L$ is a {\sl complex valued} two-component spinor.
$\hat{\Psi}^W_R$ does not interact in the standard
model because $g_V^{\nu}$=$g_A^{\nu}$. As already noted
$\hat{\Psi}^W_L$ cannot describe a Majorana particle because 
it is distinguishable from its charge conjugate
\footnote{
This fact is clearly stated in the original Lee and Yang
paper on their two-component neutrino\cite{leeyang}: 
``In this theory it is clear that the neutrino state
and the antineutrino state cannot be the same. 
A Majorana theory for such a neutrino is therefore impossible.''}.
This neutrino can obviously 
couple vectorially without becoming a four-component neutrino,
but a finite rest mass makes such a description
unavoidable.  
\\
$\bullet$  Using 
the Majorana representation we can write in general
the real and imaginary components of the Dirac equation
separately, in a way analogous to eq.(\ref{weyl}):
\begin{eqnarray}
i \gamma^{\mu} \left( {\left( {\hbar{{\partial}
\over{\partial x^{\mu}}}
+ {i k g_A^{\nu} \over c} {Z_{\mu} 
\gamma^{5}} }\right)} \hat{\Psi}_{Re} + {k g_V^{\nu} 
\over c} Z_{\mu} \hat{\Psi}_{Im} \right)
-m_{\nu} c \hat{\Psi}_{Re} = 0
\nonumber
\\
i \gamma^{\mu} \left( {\left({ \hbar{{\partial}
\over{\partial x^{\mu}}}
+ {i k g_A^{\nu} \over c} {Z_{\mu} 
\gamma^{5}} }\right)} \hat{\Psi}_{Im} - 
{k g_V^{\nu} \over c} Z_{\mu} \hat{\Psi}_{Re} \right)
-m_{\nu} c \hat{\Psi}_{Im} = 0
\label{ettore}
\end{eqnarray}
here $\hat{\Psi}=\hat{\Psi}_{Re}+i\hat{\Psi}_{Im}$.
$\hat{\Psi}_{Re}$, $\hat{\Psi}_{Im}$ are independent hermitian
operators
(see appendix 1).
The Majorana neutrino is described 
by equations (\ref{ettore}) for the case 
of vanishing vector coupling
$g_V^{\nu}$. Only in this case (and {\sl not}
for m$_{\nu}$=0) 
the two equations
decouple, and neutrinos can be fully described 
by the real part of the field $\hat{\Psi}_{Re}$
fulfilling the upper equation of (\ref{ettore}).
$\hat{\Psi}_{Re}$ is a bispinor with
four real valued components, which is
equivalent in number of independent components 
to the two complex components of the
Lee-Yang case (this is the sense in which it 
is also a ``two-component'' neutrino).
This real valued field can obviously have
a non-vanishing rest mass m$_{\nu}$
(then called ``Majorana mass'')
without becoming a four-component neutrino.
This Majorana mass might well be very different 
from the mass in the lower eq.(\ref{ettore}).
\\
A description with Weyl spinors (eq.(\ref{weyl}))
is indeed physically equivalent to a description in the Majorana
representation (eq.(\ref{ettore})) according to
Pauli's fundamental theorem. 
However it is only
the ``choice'' of the upper
equation in eq.(\ref{weyl}) which defines the
Lee-Yang neutrino. This requirement
that $\hat{\Psi}_L$ and its charge conjugate {\sl alone}
describe the observed neutrinos
is incompatible with a description
as a Majorana neutrino
also when m$_{\nu}$=0\cite{radicati} (see also the appendix 2).
\\
This {\sl paper} solves a problem in neutrino
physics but the solution
deepens the puzzle of the small neutrino masses.
\\ 
I would like to thank S.Bradbury, E.Feigl, B.Lampe,
V.E.Kuznetsov, 
P.Minkowski, S.Pezzoni, S.Raby and G.Sigl for critical 
and enlightening comments
on previous versions of this manuscript.
\\
\section{Appendices}
\subsection{The mathematical
characterization of Majorana fields as real valued fields}
Let us clarify the exact mathematical meaning
of the well known 
``real-valuedness of the wavefunction''\cite{racah}
as a defining property for Majorana particles
in a field theoretical context. 
The understanding of the Majorana field
as a field which is hermitian in the Majorana representation
is crucial for the understanding of the fundamental
difference between Lee-Yang and Majorana particles
(i.e. the difference betwen eq.(\ref{weyl}) and eq.(\ref{ettore})).
\\
The most general solution of the Dirac equation in the Majorana
representation can be written as a complete set of plane-wave 
states(see \cite{mandl} eq.(4.51)):
\begin{equation}
\hat{\Psi}(\vec{x},t) = \sum_{rk} \sqrt{m \over E (2\pi)^3}
\left( \hat{b}(k) u_r(k) e^{-ikx} + 
\hat{d}^{\dag}(k) u_r^{*}(k) e^{ikx} \right)
\label{sol}
\end{equation}
Here $\hat{b}$ and $\hat{d}$ are 
particle and antiparticle creation operator
and are given as\cite{bjorken}:
\begin{equation}
\hat{b}={1 \over \sqrt{2}} (\hat{a}_1+ i \hat{a}_2)
\\
\hat{d}={1 \over \sqrt{2}} (\hat{a}_1- i \hat{a}_2)
\end{equation}
$\hat{a}_1$ and $\hat{a}_2$ are the annihilation operators for the
Hermitian fields $\hat{\Psi}_1$ and $\hat{\Psi}_2$ 
(called $\hat{\Psi}_{Re}$ and $\hat{\Psi}_{Im}$ in eq.(\ref{ettore})) 
which are combined as $\hat{\Psi}_1$ + i $\hat{\Psi}_2$ to
obtain the most general non Hermitian field $\hat{\Psi}$.
The bispinor u${_r}$  with the  2 spin components $r$
is the usual positive energy solution of the Dirac equation.
(the superscript
W is a reminder that they are given in the Weyl representation).
k is the four momentum, m and E the particle mass and energy
respectively.
The ``supplementary condition''
for a Majorana particle  (i.e. eq.(\ref{sc}), which in the 
Majorana representation becomes : 
$\hat{\Psi}_M$=$\hat{\Psi}_M^{\dag T}$) 
requires:$\hat{b}$=$\hat{d}$. 
\\
This means a$_2$=0, i.e. $\hat{\Psi}_M$ is hermitian and
the annihilation and creation operators are
real (but not hermitian!).
The most general Majorana state can be written as:
\begin{equation}
\hat{\Psi}(\vec{x},t) = \sum_{rk} \sqrt{m \over E (2\pi)^3}
\left( \hat{a}_1(k) u_r^M(k) e^{-ikx} + 
\hat{a}_1^{\dag}(k) u_r^{M*}(k) e^{ikx} \right)
\label{sol2}
\end{equation}
This is the exact sense of the 
``reality of the wavefunction''
in the field case, in the c-number limit this 
leads to purely real wavefunctions.
\subsection{Detailed analysis of the Kayser/Shrock argument
on the neutral-current vector coupling}
Here I present a detailed proof that 
chiral left-handed states of a quantum
field (i.e. states of chirality=-1) necessarily violate
the Majorana ``supplementary condition'' (eq.(\ref{sc})).
Therefore Majorana particles cannot fulfill 
the defining condition for negative chirality states (eq.(\ref{sub})),
thus withdrawing
the basis from the Kayser/Shrock argument\cite{kaysershrock}
about the ``reappearance'' of vector interactions in Majorana
neutrino - electron scattering.
\\
Expanding in plane waves like in eq.(\ref{sol})
we can write the most
general state of negative chirality 
in the Weyl representation:
\begin{equation}
\hat{\Psi}_L^W(\vec{x},t) = \sum_{k} \sqrt{m \over E (2\pi)^3}
\left( \hat{b}(k) u_L^W(k) e^{-ikx} + 
\hat{d}^{\dag}(k) v_L^W(k) e^{ikx} \right)
\label{sol2w}
\end{equation}
here u$_r^W$ and v$_r^W$ are the usual positive and negative
energy bispinors (as always the superscript
W is a reminder that they are given in the Weyl representation).
The bispinor u$_L^W$  can by symbolized
as $\left( \begin{array}{c} 0 \\ \phi_L \end{array} \right)$,
using the Pauli two-component spinor $\phi_L$.
S$_C$ as defined in eq.(\ref{majc}) is given as
$\left( \begin{array}{cc}
0 &   i \sigma_2\\   -i \sigma_2 & 0 \\
\end{array} \right)$
in the Weyl representation. Here $\sigma_2$
is the usual Pauli matrix. A multiplication of
$\left( \begin{array}{c} 0 \\ \phi_L \end{array} \right)$
with this matrix leads to 
$\left( \begin{array}{c}i \sigma_2 \phi_L \\ 0\end{array} \right)$
which is a chiral right-handed spinor.
It can be shown\cite{bjorken}:
\begin{equation}
S_C u_L^{\dag T} = v_R ;
\\
S_C v_L^{\dag T} = u_R
\label{arma}
\end{equation}
Using eq.(\ref{arma}) to obtain  $\hat{\Psi}_L^W$
from eq.(\ref{sol2w})
it can be seen that $\hat{\Psi}_L(\vec{x},t) \neq
\hat{\Psi}_L^c(\vec{x},t)$ independent of the form of
$\hat{b}$, for each combination of individual k-components.
This means that {\it any} field with purely negative chirality
violates the Majorana condition eq.(\ref{sc}), or:
\\
{\it Majorana neutrinos cannot be in a state of pure chirality.}
\subsection{On the helicity of Majorana neutrinos}
The result of the previous section does 
{\it not} mean that Majorana fermions
cannot have a definite (e.g. left-handed) helicity! Remember
that antiparticle states with chirality=1 have 
helicity=-1\cite{sakurai}.
Consider e.g. the following
state which has a helicity of -1 in the ultra-relativistic
limit (i.e. it is ``left-handed''):
\begin{equation}
\hat{\Psi}_{h=-1} = \sum_{k} \sqrt{m \over E (2\pi)^3}
\left( \hat{a}_1(k) u_L(k) e^{-ikx} + 
\hat{a}_1^{\dag}(k) v_R(k) e^{ikx} \right)
\label{sol3}
\end{equation}
This state fulfills the Majorana condition eq.(\ref{sc}).
It describes ``a particle identical to itself''
with left handed helicity, and has 
all the properties that are  
attributed to Majoranan neutrinos 
in the standard textbooks\cite{kayser}.
It has no negative chirality however because:
\begin{equation}
{(1-\gamma^5) \over 2} \hat{\Psi}_{h=-1} = 
\sum_{k} \sqrt{m \over E (2\pi)^3}
\left( \hat{b}(k)  u_L(k) e^{-ikx} \right) \neq \hat{\Psi}_{h=-1} 
\label{sol4}
\end{equation}
Kayser and Shrock overlooked this possibility, and
erronously concluded the general validity of eq.(\ref{sub})
merely from the fact that a state has helicity=-1.
\\
The importance of the Charm II
result is, that by proving that the neutrino-electron
interaction has properties which are
directly incompatible with
the Majorana nature of the neutrino field, it provides
firm evidence that the original Lee-Yang theory, rather
than some slight modification like eq.(\ref{sol3}) describes
the physical muon neutrino.
\subsection{Reply to Comments on a previous version
of the present paper}
Finally I answer to two comments \cite{hanne,kayser97}
on a previous version of the present paper.
Hannestad\cite{hanne} accepts my argument
against Majorana neutrinos of pure chirality for
``fields''. He then argues however 
that ``states'', which are defined by the
action of a creation operator a$^{\dagger}$ on the vacuum, can
be of pure chirality.
This is impossible as I now show.
A chiral left-handed state can be created from
the vacuum state $|0>$ via:
\begin{equation}
\hat{\Psi}_L^W |0> = 
\sum_{k} \sqrt{m \over E (2\pi)^3} d^{\dagger} v_L^W  |0>= |1>_L
\end{equation}
For the charge conjugated state we have:
\begin{equation}
|1>_L^c = (\hat{\Psi}_L^W |0>)^c = 
\hat{\Psi}_L^{Wc}  |0> = 
\sum_{k} \sqrt{m \over E (2\pi)^3} b^{\dagger} 
v_R^W |0> \neq  |1>_L
\end{equation}
The last unequality holds also for the
case of a neutral particle with
$\hat{b}$=$\hat{d}$. I made the reasonable
assumption that $|0> = |0> ^c$ (to drop this assumption would
not invalidate the conclusion).
Hannestad's further discussion is similar 
to the one of Kayser and Shrock.
In fairness I have to say that in the previous version
of this paper
which Hannestad criticises I stated ``the Majorana
neutrino has to be unpolarized'', rather than 
the present more concise statement ``the Majorana 
neutrino cannot have
a definite chirality''.
\\
Kayser's recent report \cite{kayser97} mainly repeats
his arguments from Ref.\cite{kaysershrock,kayser}
in a slightly different form (as he acknowledges in his Ref.[3]).
E.g. the transition like the one from his Eq.(1) to eqs.(2) is clearly
only possible under the assumption of eq.(\ref{sub}) in
my manuscript, which does not hold for Majorana neutrinos
as explained above (see eq.(\ref{sol4})).
\\
Kayser claims that a neutral-current vector coupling of
the neutrino cannot be deduced from the Charm II results. 
In particular in the last paragraph he states that he disagrees
with the relation $g_V^{\nu e} = 2 g_V^{\nu_{\mu}} \cdot g_V^{e}$
(eq.(\ref{coup}))
which appears in Ref.\cite{charm2}, the final publication
of the Charm II collaboration on neutral
current reactions. This equation is indeed not a general
theoretical relation but is justfied in the context of
the Charm II experiment.
In the usual form of the Lagrangian for
the standard model each fermion field $\hat{\Psi}_i$
(i.e. including neutrinos)
has a vector and axial coupling constant g$_V^i$ and
g$_A^i$ (see e.g. eq.(10.1) in Ref.\cite{ppb}).
For an incident neutrino energy E$_{\nu}$ $\gg$ m$_e$
the neutral-current cross section for elastic scattering
of muon neutrinos
on electrons can then be determined from the
standard model Lagrangian  for the special
case of ``four-fermion'' problems at
center of mass energies far below
the W,Z masses, as:
\begin{equation}
{d{\sigma}\over{dy}}_{\nu,{\bar{\nu}}} = {G_F^2 m_e E_{\nu}
\over \pi} \left( (g_V^{\nu_{\mu}2} + g_A^{\nu_{\mu}2})
(g_V^{e2} + g_A^{e2}) (1+(1-y)^2)  \pm
4 g_V^{e} g_A^{e} g_V^{\nu_{\mu}} g_A^{\nu_{\mu}}
(1-(1-y)^2) \right)
\label{vivabrodo}
\end{equation}
Here and in the following equation 
the upper sign is valid for the neutrino, and the lower
sign for the antineutrino cross section. $y \equiv {E_e 
\over E_{\nu}}$ is the ratio of the kinetic energy 
of the recoil electron and the incident
$\nu_{\mu}$ or $\bar{\nu}_{\mu}$ energy.
Eq.(\ref{vivabrodo}) is similar and closely related
to the expressions for the forward-backward asymmetry
in the reaction e$^+$e$^-$ $\rightarrow$ $\ell^{+} \ell^{-}$
(eq. (10.26) of  Ref.\cite{ppb}).
The Charm II collaboration used a simplified expression
(eq.(10.17) of Ref.\cite{ppb})
to fit their data, which can be written in the following form:
\begin{equation}
{d{\sigma}\over{dy}}_{\nu,{\bar{\nu}}} =
{G_F^2 m_e E_{\nu} \over 2\pi}
\left(  (g_V^{\nu e2}+g_A^{\nu e2}) (1+(1-y)^2)  \pm
{2 g_V^{\nu e} g_A^{\nu e} (1-(1-y)^2)} \right)
\label{easy}
\end{equation}
Here ``g$_V^{\nu e}$'' and ``g$_A^{\nu e}$'' are understood
as coefficients of effective four-fermion operators.
Eq.(\ref{easy}) follows from 
eq.(\ref{vivabrodo}) if 
$g_V^{\nu e} = 2 g_V^{\nu_{\mu}} \cdot g_V^{e}$
and $g_A^{\nu e} = 2 g_A^{\nu_{\mu}} \cdot g_A^{e}$
with $g_V^{\nu_{\mu}}$=$g_A^{\nu_{\mu}}$=1/2.
Since a fit to the Charm II data to eq.(\ref{easy}) 
leads to the significant conclusion $g_V^{\nu e} \neq 0$ and
$g_A^{\nu e} \neq 0$(see section 3), it follows 
from eq.(\ref{vivabrodo}) that $g_V^{\nu_{\mu}} \neq 0$.
\\
That the Charm II neutral-current data imply the existence of
neutrino vector coupling can be seen in a very
direct way from eq.(\ref{vivabrodo}): Charm II found a small
(3.6 $\%$)
but significant (2.1 $\sigma$) difference between the total elastic
scattering cross section in the $\nu_{\mu}$-e
and $\bar{{\nu}}_{\mu}$-e case. According to eq.(\ref{vivabrodo})
this is only possible if $g_V^{\nu_{\mu}} \neq 0$.
\\
Further Kayser points out 
correctly that the Charm II collaboration did not
attempt to evaluate  $g_V^{\nu_{\mu}}$ and $g_A^{\nu_{\mu}}$
individually in Ref.\cite{charmold}.
If one takes the values for $g_V^{e}$ and $g_A^{e}$
e.g. from e$^+$ - e$^-$ experiments, it is clearly
possible in principle to obtain experimental values
for $g_V^{\nu_{\mu}}$ and  $g_A^{\nu_{\mu}}$ individually
from neutral-current scattering data,
using eq.(\ref{vivabrodo}) (except for a sign and exchange
ambiguity which already occurs in the  $g^{\nu e}$ case).
However, 
taking into account the limited precision of the Charm II data
it was a reasonable strategy to set experimental limits only
on a ``global'' neutrino coupling $g^{\nu}$ 
(which assumes $g_V^{\nu_{\mu}}$=$g_A^{\nu_{\mu}}$,
but no specific absolute value) rather than
then vector and axial constants individually.
%\pagebreak


\begin{thebibliography}{xxx}
\bibitem{majorana}
E.Majorana, Nuovo Cimento, Ser.8 {\bf 14},171 (1937).
\bibitem{bil} S.M.Bilenky, in {\it Proceedings of
the II$^{nd}$ Rencontres de Vietnam, `` At the Frontiers
of the Standard Model'', October 1995},
lanl server hep-ph/9601266 (1996).
\bibitem{bjorken}
J.D.Bjorken and  S.Drell, {\it Relativistic quantum fields}
(Mc Graw-Hill, New York, 1965).
\bibitem{mandl}
F.Mandl and G.Shaw, {\it Quantum Field Theory}
(John Wiley, Chichester, 1984).
\bibitem{sakurai} J.J.Sakurai, {\it Advanced Quantum Mechanics}
(Addison-Wesley, Reading, 1967) appendix C (Pauli's fundamental
theorem),
p.140-143(charge conjugation), p.134
(spin,momentum and velocity), p.174(CPT).
\bibitem{pauli}
W.Pauli, Rev.Mod.Phys. {\bf 13},203 (1941).
\bibitem{racah} G.Racah, Nuovo Cimento, Ser.8 {\bf 14},322 (1937).
\bibitem{furry} W.H.Furry, Phys.Rev. {\bf 54},56 (1938).
\bibitem{haberkane}
H.E.Haber and G.L.Kane, Phys.Rep.{\bf 117},75 (1985); appendix D.
\bibitem{charm2} P.Vilain et al. (Charm II collaboration),
Phys.Lett.{\bf B335},246 (1994).
\bibitem{ppb} P.M.Barnett et al., {\it Review of Particle 
Physics}, Phys.Rev.{\bf D54},1 (1996).
\bibitem{turner} E.W.Kolb and K.A.Olive, Phys.Rev. {\bf D33}, 1202
(1986).
\bibitem{kaysershrock}
B.Kayser and R.E.Shrock, Phys.Lett.{\bf 112B},137 (1982).
\bibitem{kayser}
B.Kayser,F.Gibrat-Debu and F.Perrier, {\it The Physics of
Massive Neutrinos} (World Scientific, Singapore, 1989).
\bibitem{case}
J.A.McLennan, Phys.Rev.{\bf 106},821 (1957);
\\
K.M.Case, Phys.Rev.{\bf 107},307 (1957).
\bibitem{leeyang}
T.D.Lee and C.N.Yang, Phys.Rev.{\bf 105},1671 (1957).
\bibitem{weyl} H.Weyl, ZS.f.Phys.{\bf 56},330 (1929).
\bibitem{radicati} L.A.Radicati and B.Touschek, Nuovo Cimento,
Ser.10 {\bf 5},1693 (1957).
%\bibitem{kemmer} N.Kemmer, Proc. Roy. Soc. {\bf A173},91 (1939).
\bibitem{hanne} S.Hannestad, Report no. hep-ph/9701216.
\bibitem{kayser97} B.Kayser, Report no. 
hep-ph/9703294, NSF-PT-97-1.
\bibitem{charmold} P.Vilain et al. (Charm II collaboration),
Phys.Lett. {\bf B 320},203 (1994).
\end{thebibliography}
\end{document}